\documentstyle [12pt]{article}
\begin{document}
\baselineskip .3in
\begin{titlepage}
\begin{center}{\large {\bf Anomalous Transmission in a Hierarchical Lattice}}
\vskip .2in
{\bf Anirban Chakraborti}$~^{(1)}$\\
{\it Saha Institute of Nuclear Physics},\\
{\it 1/AF Bidhan Nagar, Calcutta 700 064, India.}\\
\vskip .2in
{\bf Bibhas Bhattacharyya}$~^{(2)}$\\
{\it Department of Physics},\\
{\it Scottish Church College},\\
{\it 1 \& 3, Urquhart Square, Calcutta 700 006, India.}\\
\vskip .2in
{\bf Arunava Chakrabarti}$~^{(3)}$\\
{\it Department of Physics},\\
{\it University of Kalyani},\\
{\it Kalyani, West Bengal 741 235, India.}
\end{center}
\vskip .3in
{\centerline{\bf Abstract}}
We present an analytical method of studying ``extended" electronic
eigenstates of a diamond hierarchical lattice which may be taken
as the simplest of the hierarchical models recently proposed for
stretched polymers. We use intuitive arguments and a 
renormalization group method to determine the distribution of amplitudes of the
wavefunctions corresponding to some of these ``extended" eigenstates. An exact
analysis of the end-to-end transmission property of arbitrarily large finite 
lattices reveals  an anomalous behaviour. It is seen that while for a special
value of the energy the lattice, however large, becomes completely
transparent to an incoming electron, for the other energy eigenvalues
the transmission decreases with system size. For one such energy eigenvalue
we analytically obtain the precise scaling form of the 
transmission coefficient. The same method can easily be adopted for other
energies.
\end{titlepage}
\newpage
\noindent
{\large {\bf 1. Introduction}}
\vskip .1in

Hierarchical lattice models have played an important part in understanding the 
statistical mechanics of phase transitions. Such lattices differ,
both in topology and in geometry, from the Bravais lattices [1]. 
One major feature of such structures is their scale invariance,
which enables an exact implementation of real space renormalization group (RSRG)
techniques [1-4]. Apart from statistical mechanics, studies of the electronic
spectrum of such lattices have also revealed striking properties, not shared
by common crystalline structures [5-7]. For example, Domany et. al. [5] solved
the Schr\"{o}dinger equation on a variety of hierarchical lattices using an exact
recursive scheme. In the thermodynamic limit, the energy levels were found to
be discrete, very closely spaced and highly degenerate [5].
The electron localization problem however, is not easy to understand on 
hierarchical lattices. The fluctuating local environment around each lattice
point is likely to localize the electronic wavefunctions in most situations [5],
though for some hierarchical or fractal structures like the Seirpinski gasket and the
Vicsek fractal, ``extended" electronic states have been reported [8,9].
This aspect makes the study of electronic properties of hierarchical lattices
interesting. Normally, the ``extended" character of electronic wavefunction 
is associated with translational invariance of the underlying lattice 
and, a hierarchical lattice (like the present one) does not have any
translational periodicity by virtue of its construction (see Fig. 1). However, 
we are familiar with several one dimensional examples, viz. 
the one-dimensional random dimer [10] or quasiperiodic lattice models [11] 
where one comes across situations in which local positional correlation between
constituent `atoms' gives rise to a finite [10] or infinite [11] number of
``extended" electronic states even though these lattices are not periodic. 
In hierarchical lattices, such local positional
correlation is not always obvious. On the other hand, the topology of
the lattice in certain earlier cases has been shown to play an important
role in sustaining ``extended" electronic wavefunctions [8,9]. Such states 
sometimes exhibit anomalous transport properties, in the sense that though the 
amplitude of the wavefunction remains non-zero even at distant parts
of arbitrarily large lattices, the end-to-end transmission of any incident
wavepacket sometimes displays a power law decay [9].

Above exact results are available for hierarchical models in which 
there is a finite variety of the nearest neighbour 
environment  around a lattice site (strictly speaking, in any hierarchical lattice all  
sites are inequivalent if one looks beyond the 
nearest neighbours). The situation may thus become quite challenging if one 
tries to explore a
case where the the range of the coordination number of the lattice points increases with 
the generation of bigger and bigger lattices. One such example is the well known
diamond hierarchical lattice [1] in which the coordination number of the vertices
range from $z=2$ to $z_{max}=2^{N-1}$ for an $N^{th}$ generation lattice. 
The thermodynamics of spin models on a diamond lattice has been
studied exactly by RSRG methods [1]. But its electronic properties has not been
really studied in detail, 
a part of which we intend to address in the present communication. 
Our interest in the diamond lattice is two-fold.
Firstly, 
we wish to investigate analytically, if there exists any
``extended" electronic eigenstate though there is no translational periodicity in
this lattice.
Moreover, if such states do exist then, to our mind, it would be
interesting to study their amplitude profiles and also to study the 
transmission properties
of arbitrarily large finite lattices at those special values of the
electron-energy for which the distribution of the amplitudes is non-trivial.
Secondly, we note that similar hierarchical structures have recently been 
proposed by 
Samukhin et. al. [12] as
a possible basic structure of stretched polymers. In stretched polyacetylene
the polymer network is constructed by coupled polymer chains oriented along
some direction [12]. The $(m,n)$ hierarchical pseudolattice that serves as the
model for such polymers may be composed, according to Ref. [12], by taking
$n$  bonds forming a chain of $(n+1)$ atoms and then joining $m$ such chains in
parallel. It is then quite obvious that our diamond hierarchical
structure is a $(2,2)$ model lattice in this group (see Fig. 1). So, an analytical approach to study the
transport properties of a diamond structure seems to be an interesting step
towards the understanding of the properties of the general $(m,n)$ structure.

Recently, Zhu et. al. [13], numerically calculated the transmission coefficient 
of the $(m,n)$ polyacetylene model as
a function of electron energy for different $(m,n)$ values and at
different stage numbers.
Interestingly, they found, among very fragmented patterns, many energy values for
which the transmission coefficient turns out to be unity (or very nearly so)
for upto fifth generation structures. Also, they have reported (numerically
obtained) scaling behaviour of conductance for the $(m,n)$ structures,
though its precise form is not known.
In view of this we consider the simplest $(2,2)$ version, i.e.,
the diamond lattice and make an analytical attempt to see if such a structure
ever becomes completely or partially transparent to an incoming electron. 
Most interestingly, we find that it
is possible to get a special energy eigenvalue for which
arbitrarily large sized diamond hierarchical lattices will have
transmission coefficient equal to {\it unity}. 
We work out a
scheme based on an intuitive approach to determine the distribution of
the amplitudes of the wave
function $(\psi_i)$ for this special energy on a small sized lattice and 
then use an
RSRG approach to work out the distribution in lattices for higher generations.
The central idea is not too difficult to extend to the general $(m,n)$ case.
Using the same RSRG formalism we evaluate a whole hierarchy of 
energy eigenvalues for which we again have non-trivial distribution of the amplitudes
$\psi_i$ over the entire lattice. The end-to-end transmission across lattices
of gradually increasing size is now found to decay for this second group of eigenvalues. We
explicitly calculate the scaling behaviour for one such case and 
obtain a precise form of the power law followed by the transmission coefficient.
Scaling forms for other energies can be obtained following the scheme prescribed
by us. 
In Sec. 2  we describe our model and the method, in Sec. 3 the
transmission coefficient is discussed and we draw conclusions in Sec. 4.
\vskip .2in
\noindent
{\large {\bf 2. The Model and the Method}}
\vskip .1in

In Fig. 1, we show the first three stages of construction of the diamond lattice
following Berker and Ostlund [2]. A detailed discussion on the construction and 
topology of lattices belonging to this class is given in Ref. [1].
The atomic sites sit at every vertex of the lattice.
To study the electronic properties of the diamond lattice, we work with the
tight binding hamiltonian with nearest neighbour approximation:
\begin{equation}
H=\sum_{i} \epsilon_i \mid i~\rangle\langle ~i \mid +\sum_{<ij>}t_{ij}\mid i~\rangle\langle ~j \mid
\end{equation}
where $\epsilon_i$ is the on-site potential and $t_{ij}$ is the nearest
neighbour (n.n) hopping integral.

As we are primarily interested in the topological aspect of the system we set identical
values to all the site energies, i.e., $\epsilon_i=\epsilon$ and all the hopping integrals are 
taken to be the same, i.e. $t_{ij}=t$ (chosen to be the scale of energy). However, to facilitate the renormalization group calculations we
shall designate the site energy of a site with coordination number $z$, as
$\epsilon_z$. Therefore, in an ${N}^{th}$ generation lattice $\epsilon_i$'s 
range
from $\epsilon_2$ to $\epsilon_{z_{max}}$, where $z_{max}={2}^{N-1}$.

Let us, first of all, try to give a simple intuitive picture of how to construct a
wavefunction that will have a non-trivial disribution of amplitude over
the entire lattice, irrespective of its size. We work with the difference
equation version of the Schr\"{o}dinger equation,
\begin{equation}
(E-\epsilon_i)\psi_i= t \sum_{j\in {(n.n~ of~  i)}} \psi_j
\end{equation}
where, $\psi_j$ is the amplitude of the wave function at the $j^{th}$ site.
We shall be looking for energy eigenvalues that solve the above equation
consistently all over the lattice and yet yield a non-trivial distribution
of $\psi_j$'s.
We begin with $N=3$, i.e.,
where
we have at least one class of sites with $z>2$ ( here, $z_{max}$ is 4). 
By inspection we can
see that if we choose $E=\epsilon_2$, then this equation
satisfies the Schr\"{o}dinger equation consistently at all sites if we
demand that the amplitude of wave function vanishes at sites with
coordination number four. Extending this idea we find that indeed we can
satisfy Schr\"{o}dinger equation locally at all sites in any arbitrarily
large lattice provided we set the amplitude $\psi_i$ at a site with coordination
number $z$, equal to zero for all $z>2$. That is, in this scheme, the electron
cannot ``feel" the presence of the sites with $z>2$. This helps in attaining
an ``extended'' character of the wave function. It may be mentioned here that,
a similar technique was used earlier in the case of a Vicsek fractal [9].
Here, it is to be noted that the solutions of the Schr\"{o}dinger equation on such a lattice
will be highly degenerate and we are exhibiting
one such case only. In Figs. 2a and 2b, the amplitude distributions for
$E=\epsilon_2=0$ are given on a lattice with $N=3$ and $N=4$ respectively.
The pattern of distribution of the amplitudes for the $N=4$ case has been obtained
by joining the extremeties $A$ and $B$ (Fig. 2a) of the basic $N=3$ plaquette 
(which now becomes the building block for the next generation) 
side by side such that the vertex
$B$ of one plaquette falls on the vertex $A$ of the next (Fig. 2b). Following this
strategy we can use the $N=4$ plaquette as the basic unit and determine the
amplitude-distribution on an $N=5$ lattice by joining the $N=4$ plaquettes
at the vertices having $z=8$, where the amplitude of the wavefunction is zero.
The method can be extended easily to construct the distribution pattern for higher
generations.

The above idea can now be coupled to an RSRG scheme to extract
other possible energy eigenvalues. Starting from any
arbitrarily large finite version of the hierarchical diamond lattice, we can
renormalize the lattice $n$ times. The recursion relations for the site energies 
and hopping integral are respectively given by :
\begin{equation}
\epsilon_z(n)=\epsilon_{2z}(n-1)+\frac{2z{t}^2(n-1)}{[E-\epsilon_2(n-1)]}
\end{equation}
and
\begin{equation}
t(n)=\frac{2{t}^2(n-1)}{[E-\epsilon_2(n-1)]}
\end{equation}
Here, $\epsilon_{z}(n)$ and $t(n)$ denote respectively the 
values of the on-site potential and the hopping integral, at the $n^{th}$ stage of renormalization.

\noindent
Now, suppose we start with an $N^{th}$ generation lattice. We renormalize this
lattice $n=N-3$ times and bring it down to an ``effectively" third generation lattice
where, $z = 2$ and $4$. The site energies and the hopping integral for this
renormalized version are calculated using Eqns. $(3)$ and $(4)$. We then
apply the earlier trick on this renormalized lattice, i.e., we make the amplitudes $\psi_i$ vanish at the
$z = 4$ sites ( on the rescaled version ). This happens for
$E = \epsilon_2(n)$. This is a polynomial equation in $E$. Though, it is
difficult to provide a complete proof, yet we have performed explicit
calculations upto $N=5$ and for $n=1$ and $2$. In each case it turns out that 
the solutions of the equation
$E=\epsilon_2(n)$ satisfy the Schr\"{o}dinger equation
 all over
the lattice with a non-trivial distribution of the amplitudes of the
wavefunction. Therefore, it is tempting to make the conjecture that
the real roots of the equation $E = \epsilon_2(n)$
will correspond to the ``extended" eigenstates in the general case, if we follow
the same prescription. When mapped onto the
original lattice the wavefunction will vanish at sites with some value of $z$ 
onwards
and will
remain finite at all other lattice points with lower values of $z$.
Let us clarify this idea by discussing two specific situations. First, we 
choose $E=\epsilon_2(1)$. Now we
should have at least an $N=4$ lattice as our starting structure,
where $z_{max}=8$. We then renormalize it once to cast it into an effective
 $N=3$ - stage lattice with $z_{max}=4$. 
We find that the solutions of the equation $E=\epsilon_2(1)$ are $E=\pm2$,
with all $\epsilon_z=0$ initially and $t=1$.
Each of these energy values consistently satisfies the Schr\"{o}dinger
equation everywhere on the renormalized lattice provided we fix $\psi_i=0$
at the $z = 4$ sites
on this renormalized structure. When mapped onto the original $N=4$ - stage
hierarchical structure,
we see that the amplitude of the wavefunction vanishes only at the vertex with
the highest coordination number, i.e., 8 and it is non-zero at all other points.
The rule for constructing such a distribution may be formulated as
follows. We consider two plaquettes, type $I$ and type $II$, each being an
$N=3$ diamond lattice. In Fig. 3a, the plaquette $I$ is shown. 
In type $II$ each non-zero amplitude
of the wavefunction is of opposite sign to that at the corresponding vertex
in type $I$.
In both of them, the Schr\"{o}dinger equation is 
satisfied at each vertex for $E=2$. 
 These
two plaquettes, $I$ and $II$, are then joined end-to-end in an alternate
fashion such that the vertices with $\psi_i=0$ fall on each other (Fig. 3b). In this
way the $N=4$ structure is built, and by selecting the $N=4$ structure as
the basic unit we can construct the pattern for an $N=5$ lattice by joining
them suitably at the extreme points (with $\psi_i=0$). The scheme can go on for
other higher order lattices. It is to
be appreciated that, for higher generations $\psi_i$ will remain zero
at all vertices with $z = 8, 16, 32, ......$ .
The even number of the nearest
neighbours helps maintaining the value of $\psi_i$ equal to zero at these
sites. For the other sites with $z<4$ at any generation, the
Schr\"{o}dinger equation is locally satisfied everywhere, once it is satisfied
at a lower generation.

Second, we consider $E=\epsilon_2(2)$. Here, we need to start from an $N=5$ 
structure. We 
map this lattice onto an $N=3$ version with renormalized parameters 
and force $\psi_i=0$, as before, at the sites with $z=4$ on this renormalized lattice.
The effective $N=3$ lattice with the distribution of the amplitudes is
presented in Fig. 4a. When unfolded to retrieve the original $N=5$ lattice,
we now find that the amplitude is zero only at the vertices with $z=16$, and
three new amplitudes $\pm{a}$, $\pm{b}$ and $\pm{c}$ appear at $z=2$ and $4$ - 
sites. One quarter of the full $N=5$ lattice is shown in Fig. 4b with the 
distribution of the values $a$, $b$ and $c$. The complementary portion with
$-a$, $-b$ and $-c$ is not shown, but can easily be conceived of. 
In order to satisfy the Schr\"{o}dinger equation consistently at each 
vertex, we see that $a$, $b$ and $c$ should take values 
$(E^2-8)/8E$, $(E^2-8)/8$ and $E/8$ respectively, provided the energy $E$ is a
solution of the equation
$E^4-12E^2+16=0$. This is precisely the polynomial equation that
is obtained by setting $E=\epsilon_2(2)$. We thus confirm that for every root
of this equation we are able to construct extended eigenstates 
consistent with the Schr\"{o}dinger equation. 
We expect that, reasoning in the same manner as in the above cases,
we are likely to uncover a whole set of eigenvalues by solving the equation
$E=\epsilon_2(n)$ with $n=1$, $2$, $3$, ..., 
in an arbitrarily large lattice for which the wavefunction
will be ``extended" in the sense described earlier.

Before ending this section we point out two other possibilities
of getting ``extended" type
of states:

{\it (i)} We consider the third generation lattice, where $z_{max}=4$. It
is quite obvious that by choosing $E=\epsilon_4$ (which happens to be
equal to $\epsilon_2$ in the bare length scale in our model),
we can have a consistent
solution of the Schr\"{o}dinger equation on this lattice in which the amplitude 
of the wavefunction is zero on all vertices with $z=2$ and alternates between 
$\pm1$ on the vertices with $z=4$. This is, of course, one of the possible 
configurations. But we can carry on the process of construction for other
eigenvalues by setting
$E=\epsilon_4(n)$ for an $N=n+3$ generation lattice,
and by demanding that $\psi_i$ vanishes identically on all 
$z=2$ vertices on the $n$ step renormalized version of the same lattice. In Fig. 5, we 
exhibit the distribution of amplitudes for $E=\epsilon_4(1)$ on a
fourth generation lattice. Distribution for higher generations and other
eigenvalues can be obtained by extending the earlier ideas.
Similar results can be obtained for the general case with $E=\epsilon_{2z}(n)$. 

{\it (ii)} The other possibility refers to a specific initial 
choice of the site energies.
By looking at the recursion relations $(3)$ and $(4)$ for the on-site term and hopping integral respectively, we find
that if we start with a model where $\epsilon_{2z}=\epsilon_z - zt$,
then for $E=\epsilon_2 + 2t$, each on-site term and the hopping integral
exhibit a fixed point behaviour. It implies that the nearest neighbour
hopping integral does not flow to zero under iteration and we have a
non-vanishing `connection' between nearest neighbouring sites at all
length scales. This is a clear signature of the corresponding eigenstate
being extended [14].

We will now describe how to investigate the transmission characteristics
of a diamond hierarchical lattice. We will emphasize on the analytical treatment of
the recursion relations and will discuss the behaviour of the transmission
coefficient $T(E)$ for the two cases with $E = \epsilon_2$ and 
$E = \epsilon_2(1)$ respectively, for which the transmission coefficient displays
totally opposite characteristics. 
$T(E)$ for the other energies $( E=\epsilon_{2z}(n) )$ can be obtained by following the method
adopted for these cases.
\vskip .2in
\noindent
{\large{\bf 3. Analysis of the Transmission Coefficient}}
\vskip .1in

For calculating the transmission coefficient $T(E)$, we attach two semi-infinite
perfectly ordered leads to the two `diametrically' opposite vertices having
the maximum coordination number in any generation $N$. The original lattice is 
then
renormalized $n( =N-2 )$ times, so that we are left with a basic $N=2$ rhombus 
with
 four vertices
each having an effective site energy $\epsilon_2(n)$ and the nearest neighbour
hopping integtral $t(n)$ [Fig. 6]. This elementary rhombus is now folded into a
`dimer' with site energy and nearest neighbour hopping integral respectively 
given by
\begin{displaymath}
{\tilde{\epsilon}}=\epsilon_2(n)+\frac{2t^2(n)}{[E-\epsilon_2(n)]}
\end{displaymath}
and
\begin{displaymath}
{\tilde{t}}=\frac{2t^2(n)}{[E-\epsilon_2(n)]}
\end{displaymath}

Following the standard procedure [15] it is then easy to show that
\begin{equation}
 T(E)=\frac{4~{\sin}^2k}{{[P_{21}-P_{12}+(P_{22}-P_{11})\cos k]}^2+{(P_{11}+P_{22})}^2{\sin}^2k}
\end{equation}
where the elements of the matrix $P$ are:
\begin{equation}
P_{11}=\left[\left(\frac{E-\tilde{\epsilon}}{\tilde{t}}\right)^2-1\right]\frac{\tilde{t}}{t_0}
\end{equation}
\begin{equation}
P_{12}=-\left(\frac{E-\tilde{\epsilon}}{\tilde{t}}\right)=-P_{21}
\end{equation}
and
\begin{equation}
P_{22}=-\frac{t_0}{\tilde{t}}
\end{equation}

Here, $k = \cos^{-1}[(E-\epsilon_0)/2t_0]$, where, $\epsilon_0$ and $t_0$ refer
to the on-site potential and the hopping integral respectively, of the ordered
lead.

In order to understand the behaviour of $T(E)$ for large systems at any
particular energy, we must analyze how the matrix elements $P_{ij}$ behave
for large number of RG iterations, $n$. As a lattice of any generation  
should finally be reduced to a basic $N=2$ rhombus having only
$\epsilon_2(n)$ and $t(n)$ (see Fig. 5), we must analyse the flow patterns
of $\epsilon_2$ and $t$ under successive RSRG iterations. The evolution of
these two parameters ultimately controls $\tilde\epsilon$ and $\tilde t$, and
hence the matrix elements $P_{ij}$.
Let us discuss it for two specific
cases. Throughout the analysis we will set all $\epsilon_z=0$ and $t=1$.

{\it Case (i)} $E = \epsilon_2 =0$

From direct calculations we find that as $E \rightarrow 0$, the leading
behaviours (in $E$) of $t$ and $\epsilon_2$ are given by,
\begin{equation}
t(n)\sim\frac{2t^2}{E}
\end{equation}
\begin{equation}
\epsilon_2(n)\sim\frac{-4t^2}{E}
\end{equation}
for $n=2, 3, 4$ and $5$.
We thus assume  these forms to be true for any
arbitrary value of $n$, viz., $n=m$. Then proceeding according to
the standard method of induction, we can prove, using the recursion
relations $(3)$ and $(4)$, that Eqns. $(9)$ and $(10)$ indeed hold good for
$n=m+1$ as well. Therefore, we accept the above forms as the leading terms
in the expressions for $t(n)$ and $\epsilon_2(n)$ for $E \rightarrow 0$ and
for any arbitrary value of $n$ with $n\geq{2}$. It is now easy to work out
an expression for $(E-\tilde{\epsilon})/\tilde{t}$, which is given, to the
leading order in $E$, by
\begin{equation}
\frac{E-\tilde{\epsilon}}{\tilde t} = \frac{E^2}{2t^2} + 1
\end{equation}
A direct subtitution of the above result in the expressions of $P_{ij}$
shows that, $P_{11}=P_{22} \rightarrow 0$ and $P_{12}=-P_{21}=-1$ as $E \rightarrow 0$.
The expression for $T(E=0)$ now becomes :
\begin{equation}
T(E=0)=\frac{4~{\sin}^2k}{{[P_{21}-P_{12}]}^2}={\sin}^2k=1-\frac{\epsilon_0^2}{4t_0^2}
\end{equation}

If we select $\epsilon_0=0$, then $T(E=0)$ is unity, and any arbitrarily large 
diamond hierarchical lattice 
becomes completely transparent to an incoming electron with $E=0$.

{\it Case (ii)} $E=\epsilon_2(1)=2$

The central idea of the analysis for case $(i)$ can now easily be extended to study
the general situation, where $E=\epsilon_2(n)$. The results, however, turn
out to be totally different, as we have checked numerically by solving the
equation $E-\epsilon_2(n) = 0$ for several values
of $n$. We present below analytical results for $E=\epsilon_2(1)=2$ for steps
$n\geq 2$.
Once again, we observe the behaviour of $t$ and $\epsilon_2$
around $E=2$ for successive iterations. We set $E=2+\delta$, $\delta$ being
infinitesimally small, and find, by direct calculation that, for $n\geq 2$
\begin{equation}
\epsilon_2(n) = \frac{2}{\delta} + f_n + \cal{O}(\delta)
\end{equation}
\begin{equation}
t(n) = -\frac{1}{\delta} + g_n + \cal{O}(\delta)
\end{equation}
upto leading order in $\delta$, where $f_n$ and $g_n$ are respectively given by 
\begin{displaymath}
f_n = \frac{2^{2n}}{9}+\frac{2n}{3}-\frac{11}{18} 
\end{displaymath}
\\
and
\begin{displaymath}
g_n = \frac{2^{2n-2}}{9}-\frac{n}{3}+\frac{35}{36}
\end{displaymath}\\
The above forms set in, as in case {\it (i)}, after the first iteration, and
hold perfectly well for $n=2,3,4$ and $5$. We now make use of the recursion
relations (3) and (4) to find that the result is true for $(n+1)^{th}$ stage as
well. Thus, we take Eqns. (13) and (14) to represent the general $n$ behaviour
of $\epsilon_2(n)$ and $t(n)$ for any $n \geq 2$. The leading $\delta$ - 
behaviour of $\tilde\epsilon$ and $\tilde t$ are obtained to be,
\begin{displaymath}
\tilde\epsilon = \frac{1}{\delta}+\frac{f_{n+1}}{2} 
\end{displaymath}\\
and
\begin{displaymath}
\tilde{t} = -\frac{1}{\delta}+g_{n+1}
\end{displaymath}\\
respectively, which lead to the equation
\begin{equation}
\frac{E-\tilde\epsilon}{\tilde t} = \frac{2^{2n}}{3}-\frac{4}{3}
\end{equation}
However, it should be noted that in order that the present analysis is valid, we
must have a finite (although large) number of iterations $n$ such that $f_n$ and
$g_n$ are also finite quantities and are small compared to $1/ \delta$ as 
$\delta \rightarrow 0$.
The matrix elements now read 
(neglecting terms of the order of $\delta^2$ and taking the
limit $\delta \rightarrow 0$),
$P_{11}=-2(4^n-4)/3$ 
and, $P_{22}=0$,
 $P_{12}=-P_{21}=-1$. For large (but finite) values of $n$ one can now show, using
Eqn. (5) that
\begin{displaymath}
T_n(E=2) \sim 2^{-4n}
\end{displaymath}

It is quite obvious that, depending on the value of the energy, one has
to select the on-site term and the hopping integral for the lead suitably,
so that the energy does not fall beyond the ``band'' of the ordered
lead or even coincide with the ``band-edge". In both 
cases the use of the expression (5) will not be meaningful.

We have also numerically calculated
$T(E)$ for $E=\epsilon_2(2)$, $E=\epsilon_2(3)$ and
$E=\epsilon_4(1)$ for lattices starting from
$N=3$ upto $N=6$. We observe a gradual attenuation in the value of the transmission
coefficient according to our expectation. The present results thus provide an example of 
what we may call an
``atypical" extended state [9] where, though the wavefunction displays
non-zero amplitudes even at the farthest portions of an arbitrarily large lattice, the end-to-end
transmission decays with increasing lattice size. We expect similar
behaviour of $T(E)$ for other cases (with $E=\epsilon_{2z}(1)$) also.

\noindent
{\it The fixed point behaviour of T(E) :}

Before ending this section, we discuss the case where we have a fixed-point
behaviour of the hamiltonian parameters. For this, we take a model
with $\epsilon_{2z}=\epsilon_z-zt$ and set $E=\epsilon_2+2t$. All parameters
then remain unaltered under RSRG and we find that for $\epsilon_2=0$ and $t=1$,
the numerical value of the transmission coefficient is given by, 
\begin{equation}
T(t_0) = \frac{4(t_{0}^2-1)}{t_{0}^4}
\end{equation}
where we have set the site energy of the lead $\epsilon_0$, equal to zero. Naturally,
we have to choose any suitable value for $t_0$ so that the above energy
remains within the ``band" of the ordered lead.
The above expression for $T$ remains fixed for arbitrarily large versions of
a diamond lattice. The wave function is definitely extended as $t$ does not
flow to zero under RSRG.
\vskip .2in
\noindent
{\large{\bf 4. Conclusion}}
\vskip .1in

We have presented  a hierarchical lattice model where the coordination
number of the lattice points range from $2$ to $2^{N-1}$ depending on the
generation index $N$. In such a lattice there exist ``extended" type of
electronic states some of which have been identified and the corresponding
eigenvalues have been calculated using renormalization group ideas. We
also presented an exact analysis of the end-to-end transmission coefficient
to reveal that the lattice, irrespective of its size, becomes completely
transparent to an electron with energy $E=0$, while for other energies
the transmission coefficient has a scaling behaviour. We obtained an exact
form of the scaling for a specific energy, and the other forms can be
obtained using the same methodology, though we did not present the other
analytical results here.
\pagebreak

\noindent
{\large \bf References}
\vskip .1in

\noindent
{\it e-mail addresses} :

\noindent
$^{(1)}$ papluchakrabarti@hotmail.com

\noindent
$^{(2)}$ bibhas@cmp.saha.ernet.in

\noindent
$^{(3)}$ arunava@klyuniv.ernet.in and rkm@cmp.saha.ernet.in
\vskip .1in

\noindent
[1] R. B. Griffiths and M. Kaufman, Phys. Rev. {\bf B 26}, 5022 (1982).

\noindent
[2] A. N. Berker and S. Ostlund, J. Phys. {\bf C 45}, 4961 (1979).

\noindent
[3] Y. Gefen, B. Mandelbrot and A. Aharony, Phys. Rev. Lett. {\bf 45}, 855 (1980).

\noindent
[4] S. Alexander and R. Orbach, J. Phys. Lett. {\bf 43}, L625 (1982);
    J. R. Banavar and M. Cieplak, Phys. Rev. {\bf B 28}, 3813 (1983).

\noindent
[5] E. Domany, S. Alexander, D. Bensimon and L. P. Kadanoff,
Phys. Rev. {\bf B 48}, 3110 (1983).

\noindent
[6] R. Rammal and G. Toulouse, Phys. Rev. {\bf B 49} ,1194 (1982).

\noindent
[7] W. Schwalm and M. Schwalm, Phys. Rev. {\bf B 39}, 12872 (1989);
W. Schwalm and M. Schwalm, Phys. Rev. {\bf B 47}, 7848 (1993);

\noindent
[8] X. R. Wang, Phys. Rev. {\bf B 51}, 9310 (1995) ;
A. Chakrabarti, J. Phys.:Condens. Matter {\bf 8}, 10951 (1996).

\noindent
[9] A. Chakrabarti and B. Bhattacharyya , Phys. Rev. {\bf B 54}, R12625 (1996);
    A. Chakrabarti, J. Phys.: Condens. Matter {\bf 8}, L99 (1996). 

\noindent
[10] D. H. Dunlap, H. -L. Wu and P. Phillips, Phys. Rev. Lett. {\bf 65}, 88 (1990).

\noindent
[11] A. Chakrabarti, S. N. Karmakar and R. K. Moitra, Phys. Rev. {\bf B 50} ,
13276 (1994).

\noindent
[12] A. N. Samukhin, V. N. Prigodin and L. Jastrabik, Phys. Rev. Lett. {\bf 78},
326 (1997).

\noindent
[13] C. P. Zhu, S. J. Xiong and T. Chen, Phys. Rev. {\bf B 52}, 12848 (1998).

\noindent
[14] B. W. Southern, A. A. Kumar, P. D. Loly and A. -M. S. Tremblay,
Phys. Rev. {\bf B 27}, 1405 (1983).

\noindent
[15] A. Douglas Stone, J. D. Joannopoulos and D. J. Chadi,
Phys. Rev. {\bf B 62}, 5583 (1981).

\newpage
\noindent
{\large \bf Figure Captions}
\vskip .1in

\noindent
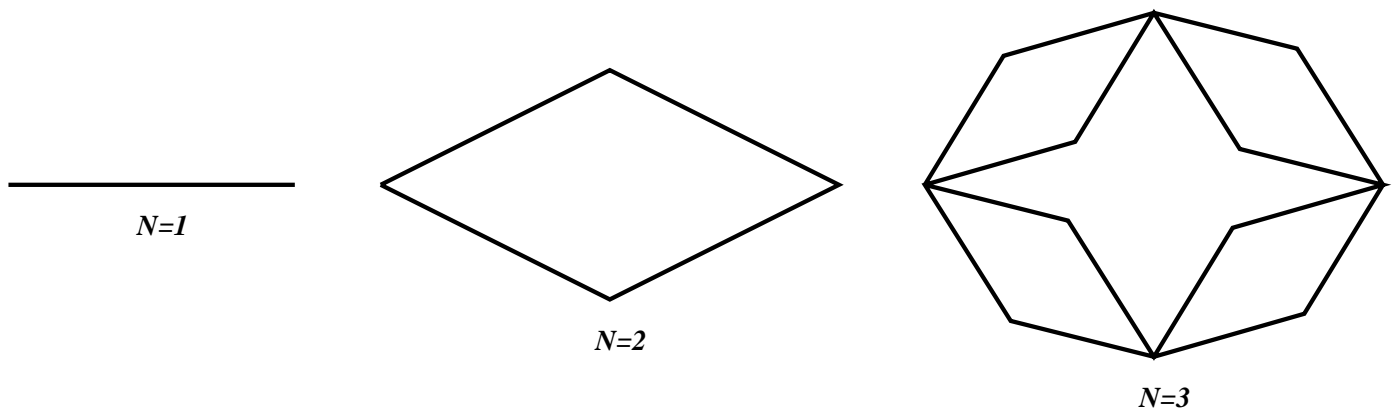
{\bf Fig. 1}:  First three stages (i.e., $N=1,2,3$) of the construction of a diamond
hierarchical lattice.

\noindent
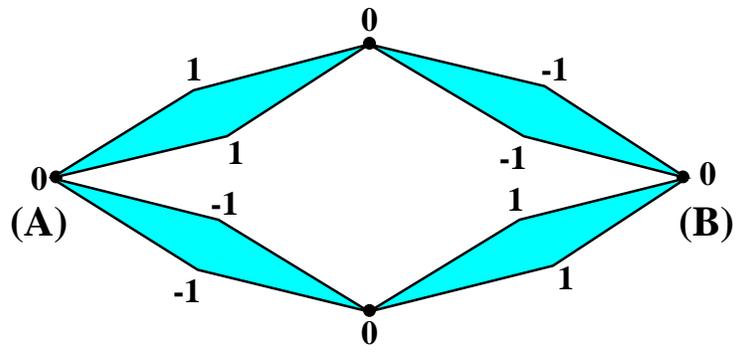
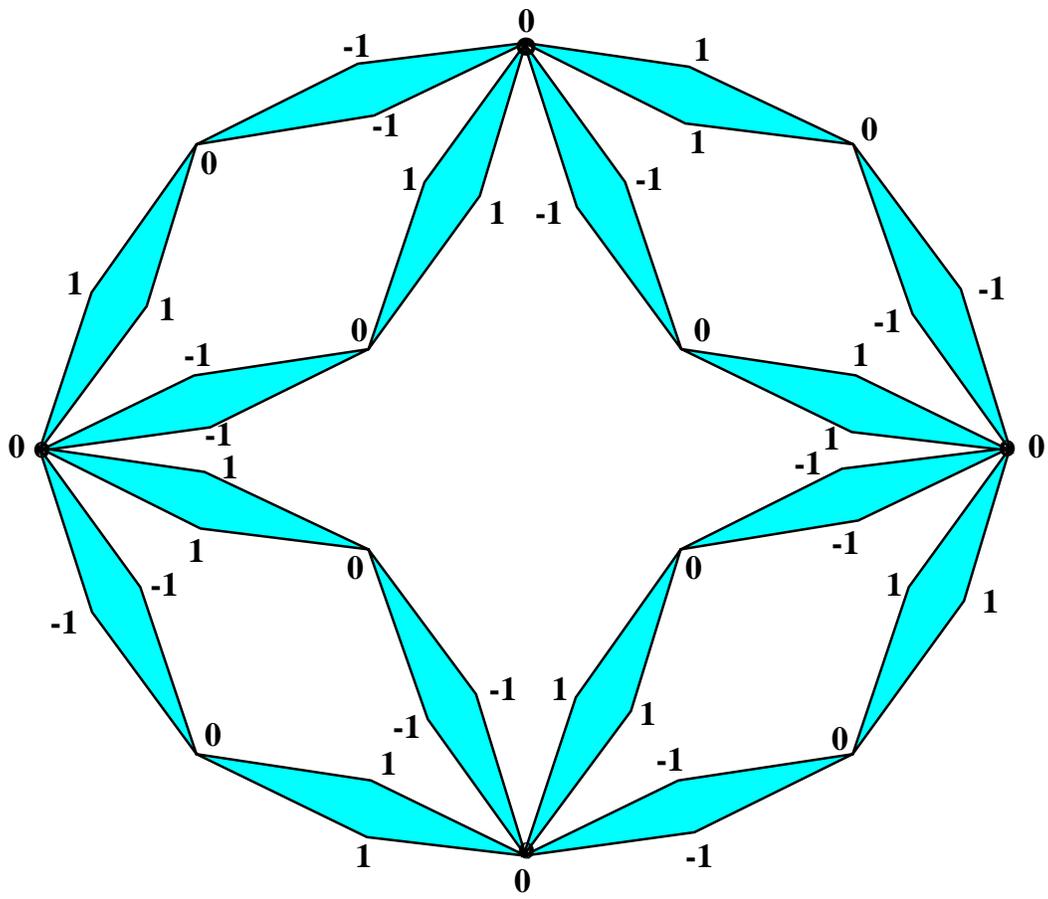
{\bf Fig. 2}: Distribution of the amplitudes of an extended
wavefunction at $E=0$ on {\bf (a)} an $N=3$ lattice and {\bf (b)} an $N=4$ lattice. 
All $\epsilon_i$'s have been set equal to zero and $t=1$. $\psi_i$'s take on
values $-1$, $0$ and $1$ on different sites ($i$).

\noindent
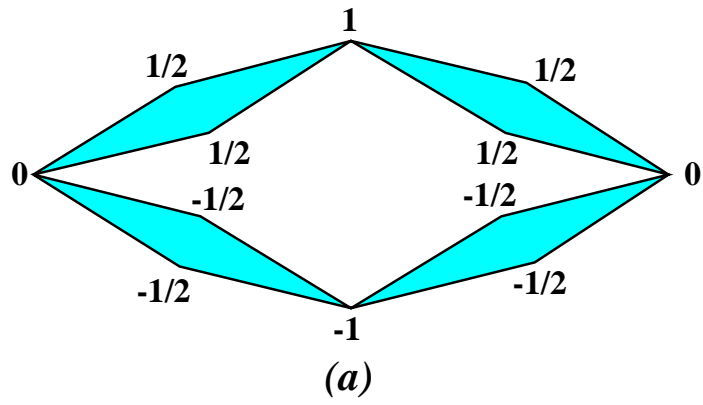
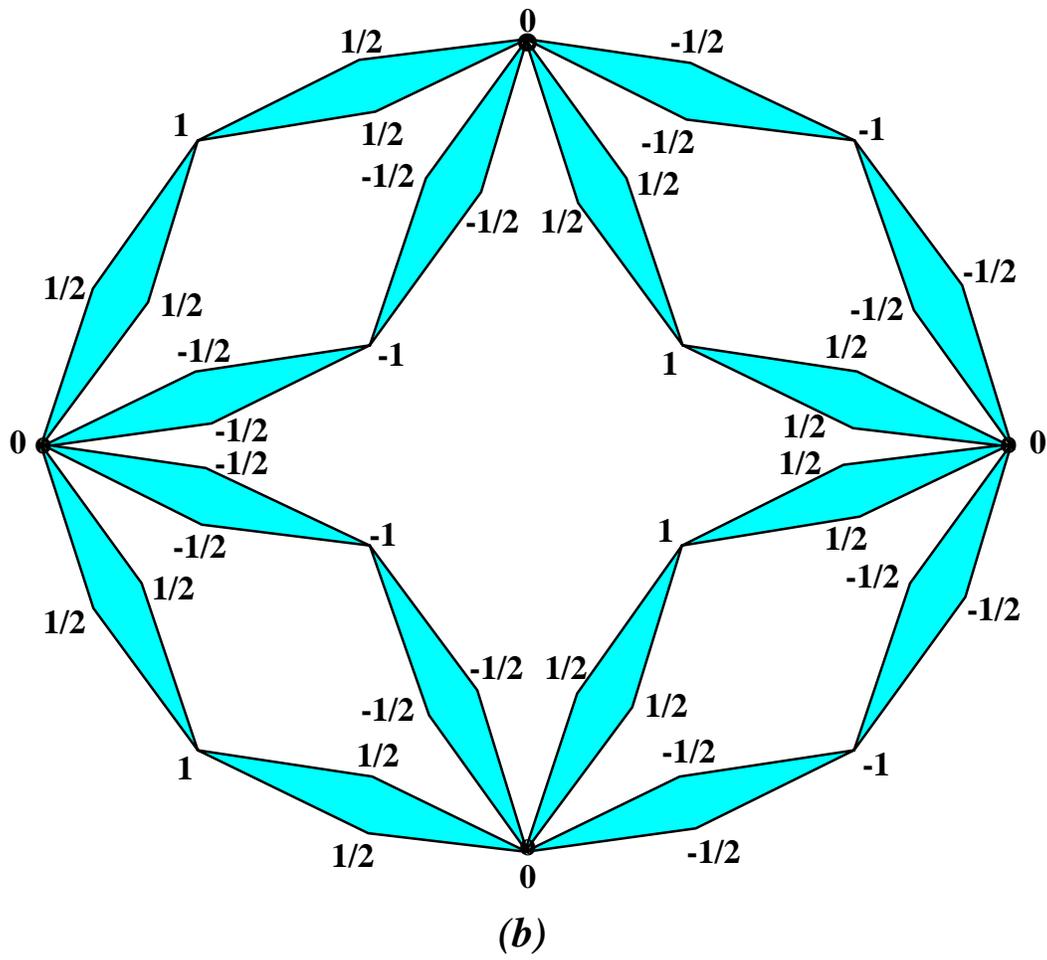
{\bf Fig. 3}:  Amplitudes of a wavefunction on {\bf (a)}
the basic plaquette $I$ which acts as a building
block of an $N=4$ lattice and {\bf (b)} an $N=4$ lattice for
$E=\epsilon_2(1)$.
All $\epsilon_i$'s have been set equal to zero and $t=1$. $\psi_i$'s take on
values $-1$, $-1/2$, $0$, $1/2$ and $1$ on different sites ($i$).

\noindent
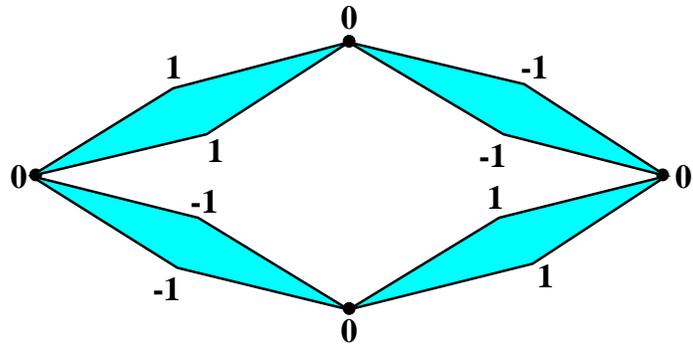
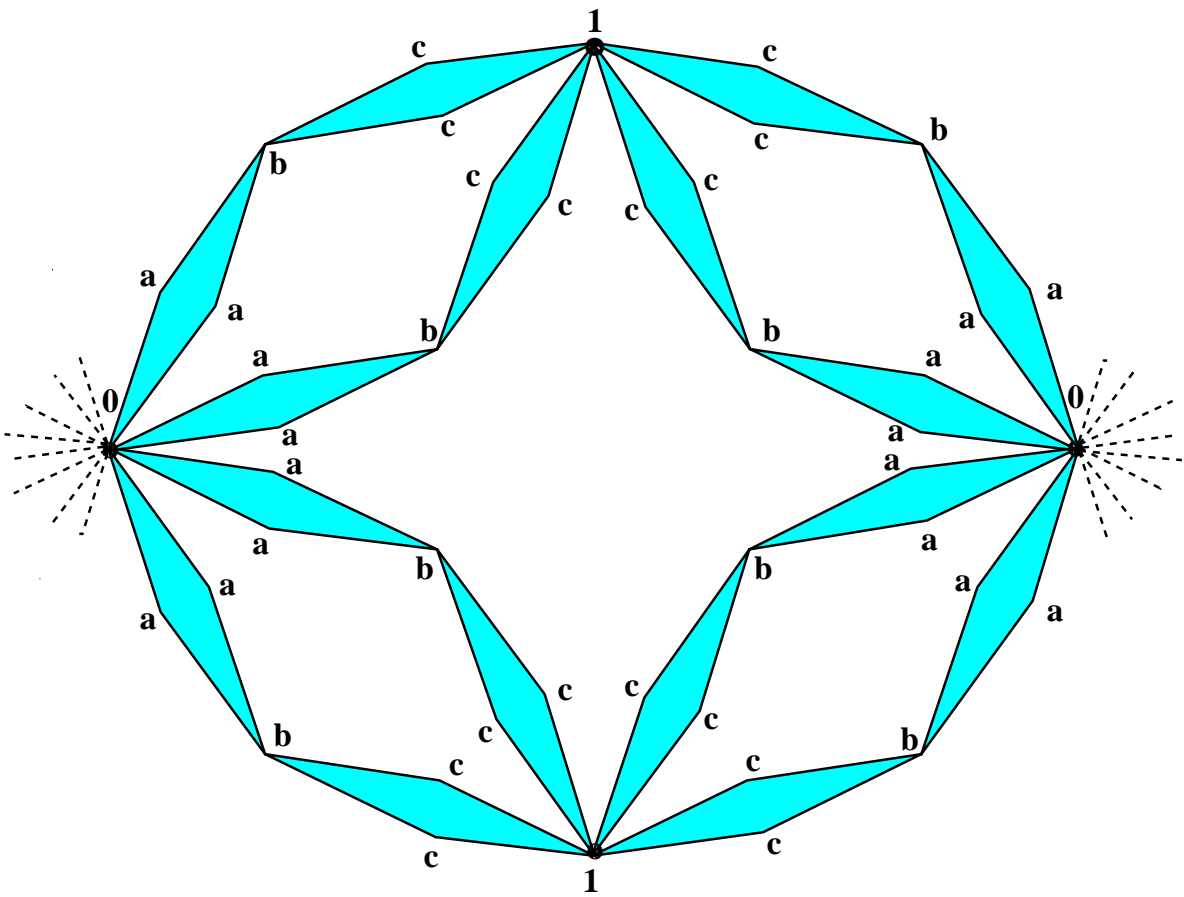
{\bf Fig. 4}: Amplitude distribution  on {\bf (a)} an effectively $N=3$
plaquette obtained by renormalizing an $N=5$ lattice twice and
{\bf (b)} one quarter of the original $N=5$ version for $E=\epsilon_2(2)$.
Dashed lines at the two extreme vertices indicate the presence of complementary
plaquettes.
All $\epsilon_i$'s have been set equal to zero and $t=1$. The values of $a$, $b$
and $c$ are given in the text.

\noindent
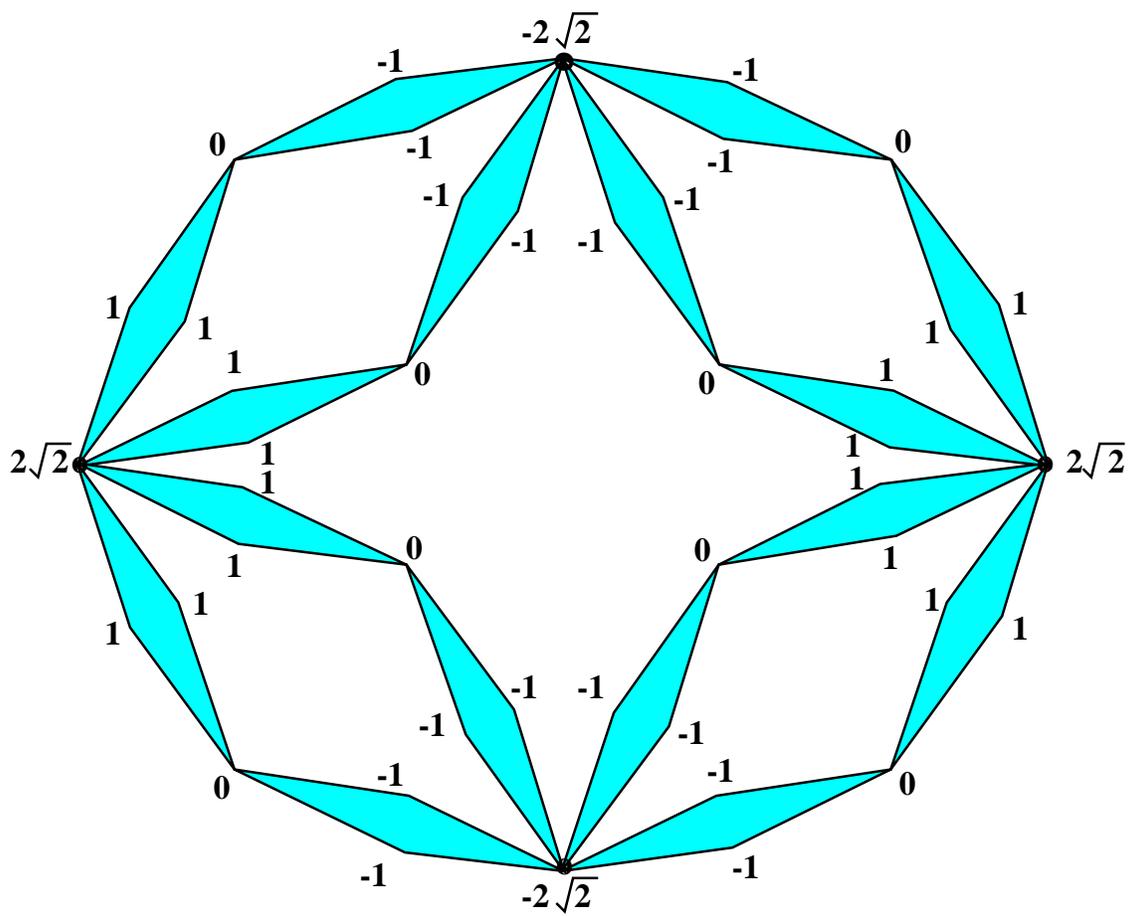
{\bf Fig. 5}:  Amplitudes of a wavefunction for $E=\epsilon_4(1)$ on
an $N=4$ lattice. 
All $\epsilon_i$'s have been set equal to zero and $t=1$. $\psi_i$'s take on
values $-2\sqrt{2}$, $-1$, $0$, $1$ and $2\sqrt{2}$ on different sites ($i$).

\noindent
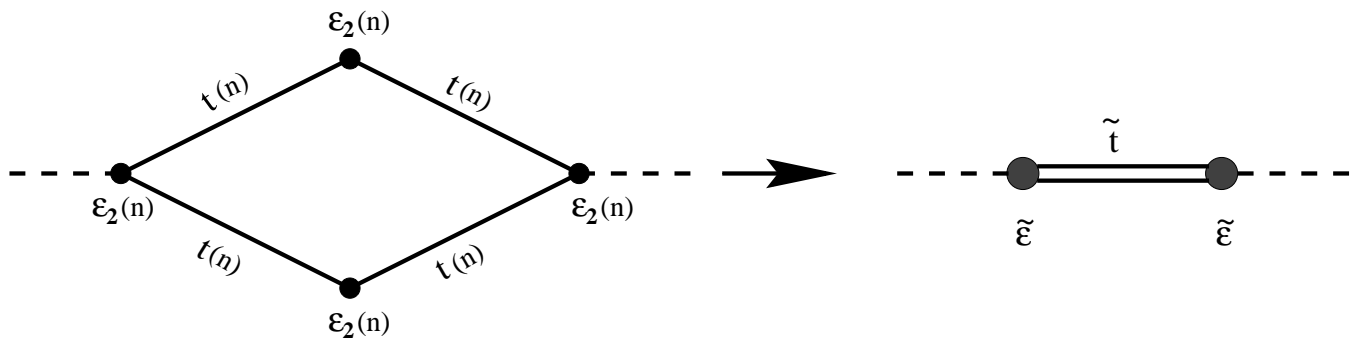
{\bf Fig. 6}:  Reduction of an $n$ times renormalized lattice to an
effective dimer. The leads are shown as dashed lines.
\end{document}